\documentclass[a4paper,twocolumn,accepted=2021-07-22]{quantumarticle}

\pdfoutput=1
\usepackage[utf8]{inputenc}
\usepackage[english]{babel}
\usepackage[T1]{fontenc}
\usepackage{csquotes}
\usepackage{physics}

\usepackage[dvipsnames]{xcolor}

\usepackage[colorlinks=true]{hyperref}

\usepackage[sorting=none,backend=bibtex]{biblatex}
\usepackage{amsmath,amsfonts}
\DeclareMathOperator*{\argmin}{arg\,min}
\DeclareMathOperator*{\argmax}{arg\,max}

\usepackage{listings}
\usepackage{parskip}

\usepackage{graphicx}
\graphicspath{{images/}}

\begin{document}
\title{An efficient quantum algorithm for the time evolution of parameterized circuits}

\author{Stefano Barison}
\affiliation{Institute of Physics, \'{E}cole Polytechnique F\'{e}d\'{e}rale de Lausanne (EPFL), CH-1015 Lausanne, Switzerland}
\author{Filippo Vicentini}
\affiliation{Institute of Physics, \'{E}cole Polytechnique F\'{e}d\'{e}rale de Lausanne (EPFL), CH-1015 Lausanne, Switzerland}

\author{Giuseppe Carleo}
\affiliation{Institute of Physics, \'{E}cole Polytechnique F\'{e}d\'{e}rale de Lausanne (EPFL), CH-1015 Lausanne, Switzerland}

\begin{abstract}
We introduce a novel hybrid algorithm to simulate the real-time evolution of quantum systems using parameterized quantum circuits.
The method, named "projected - Variational Quantum Dynamics" (p-VQD) realizes an iterative, global projection of the exact time evolution onto the parameterized manifold. In the small time step limit, this is equivalent to the McLachlan's variational principle. Our approach is efficient in the sense that it exhibits an optimal linear scaling with the total number of variational parameters. Furthermore, it is global in the sense that it uses the variational principle to optimize all parameters at once. The global nature of our approach then significantly extends the scope of existing efficient variational methods, that instead typically rely on the iterative optimisation of a restricted subset of variational parameters.   
Through numerical experiments, we also show that our approach is particularly advantageous over existing global optimisation algorithms based on the time-dependent variational principle that, due to a demanding quadratic scaling with parameter numbers, are unsuitable for large parameterized quantum circuits.
\end{abstract}

\maketitle

\section{Introduction}
\label{sec:intro}

In recent years, our ability to manipulate and measure quantum systems has improved tremendously.
Among all experimental platforms, the number of addressable qubits has increased remarkably: both Google \cite{Arute19Nat} and IBM \cite{Sager20PRR} have reported superconducting circuits chips with $\approx 50$ qubits and according to their public timelines they expect to build 1000-qubit devices within the next 2 years.
Despite this impressive development, universal quantum-computing remains still far in the future. Algorithms such as Shor's \cite{Shor94Book}, Quantum-Fourier Transform \cite{Coppersmith94QFT} or general Quantum Simulators \cite{Santoro2006,Kassal2008,georgescu14qs}
 require a number of operations (gates) at least polynomial in the qubit number. 
However, in the absence of large-scale quantum error correction, the number of gates that can be applied is at present strongly limited by hardware noise and decoherence.

To circumvent the problem, current generations of quantum algorithms rely on a hybrid classical-quantum approach \cite{Peruzzo_2014,Ying2017rte,ollitrault2019quantum,Motta_2019,cerezo2020variational,Beckey20}.
A typical example is given by the classical optimisation of a quantum circuit encoding the solution of a given problem.
Quantum circuits can be used as the model of a machine-learning problem \cite{Biamonte_2017,borujeni2020quantum}, such as a quantum classifiers \cite{Romero17QST,Kerenidis19NIPS}, quantum kernel machines \cite{Schuld19PRL,Schuld20PRA,Havlek19Nat}, quantum Boltzmann machines \cite{Amin18PRX} or convolutional networks \cite{Cong19NatPhys} to name a few. 
Quantum circuits can also be used to directly approximate the state of an interacting quantum system \cite{OMalley16PRX,Kandala2017,Bauer20ChemRev}. 
In particular, several of the proposed hybrid algorithms extend the use of the variational method to quantum computing: a trial quantum state (ansatz state) with a tractable number of parameters and mild circuit depth is considered, then these parameters are optimized in order to approximate a target state as accurately as possible.

For quantum simulation, several variational, resource friendly alternatives to Trotterization \cite{Trotter1959,Suzuki1991,abrams97qs,Ortiz2001} have been proposed \cite{Ying2017rte,Yuan2019tv,C_rstoiu_2020,commeau2020variational,bharti2020quantum}.
Among those, an interesting approach introduced in 2017 \cite{Ying2017rte} is the The Time-Dependent Variational Algorithm (TDVA).
This method, based on a reformulation of the Dirac-Frenkel and McLachlan variational principle \cite{dirac1930,frenkel1934,McLachlan1964}, encodes the state into a variational circuit and iteratively updates the parameters by solving a Euler-Lagrange equation of motion. 
The approach is conceptually close to the study of quantum dynamics using classically parameterized many-body wave functions \cite{HaegemanPRL11,HaegemanPRB16,carleo_localization_2012,carleo_light-cone_2014}.
At variance with the classical many-body counterparts, the computational cost of the TDVA approach is dominated by the stochastic estimation of the Quantum Geometric Tensor \cite{Kolodrubetz17qgt,Bukov19qgt} and its inversion, resulting in an expensive quadratic cost in the total number of variational parameters.
To alleviate this issue, new variational methods based on partial, local optimisations of the variational parameters have been recently proposed \cite{benedetti2020hardwareefficient,slattery2021unitary,Barratt_2021}.

In this paper we propose a hybrid quantum algorithm for approximating the real time evolution of an interacting quantum system, called projected-Variational Quantum Dynamics (p-VQD). 
This algorithm is both global -- it optimizes all parameters at once -- and efficient -- it scales linearly with the number of parameters. 
Moreover, it does not require auxiliary (ancilla) qubits and the depth of the circuit is constant throughout all the simulation.
The structure of this paper is as follows: in Section \ref{sec:methods} we present the algorithm in order to simulate real time evolution of quantum systems, while in Section \ref{sec:results} we benchmark our algorithm on a Transverse Field Ising model, assessing the accuracy of the method on an ideal quantum simulator performing single and multi shot executions. We conclude, in Section \ref{sec:discussion}, with some considerations on the proposed algorithm.

\begin{figure*}
\begin{centering}
        \includegraphics[width=1.65\columnwidth]{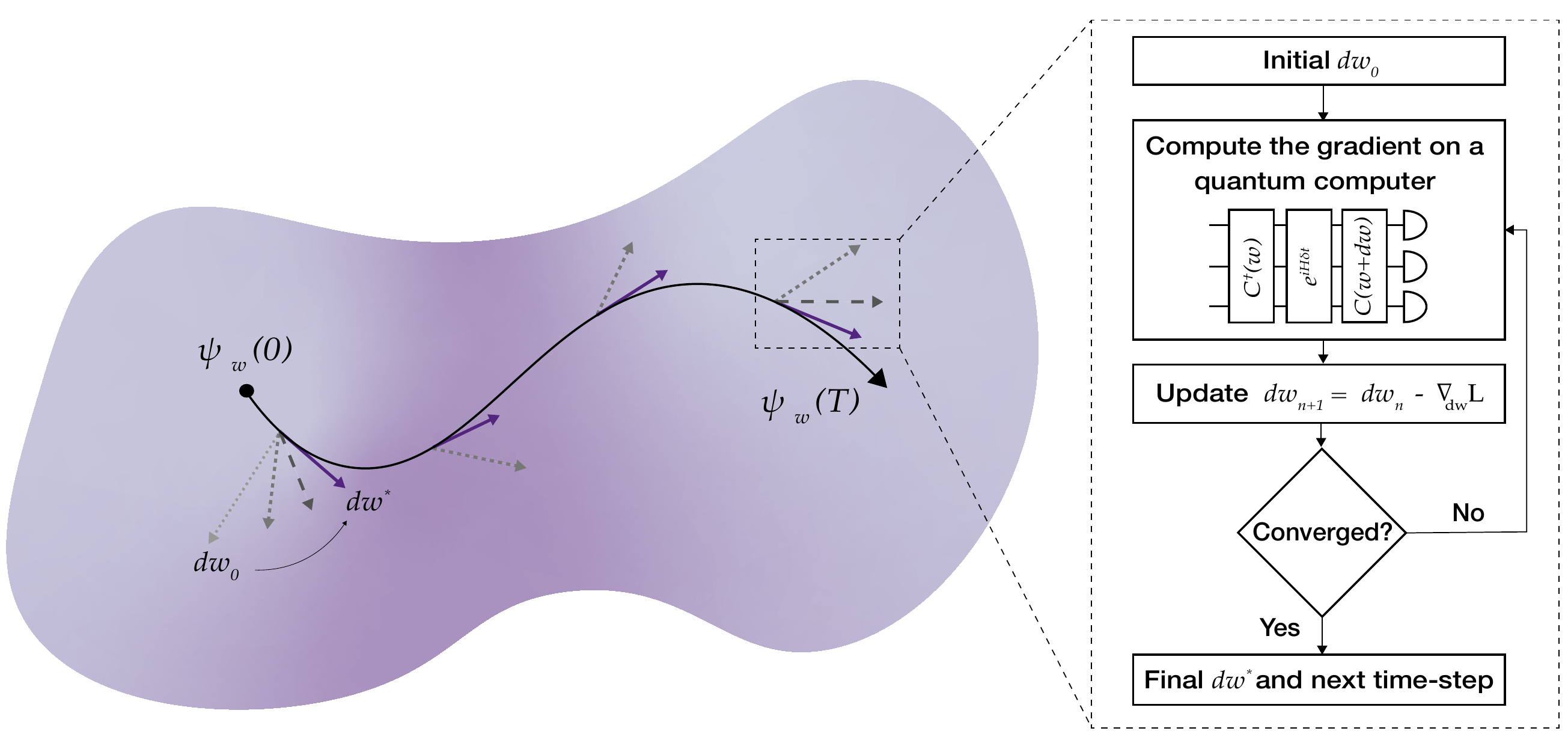} 
        \caption{Sketch of the p-VQD algorithm. We follow the real time evolution of the ansatz state $\psi_{w}(t)$ in the Hilbert space by optimizing the parameter variation $dw$ at every time step. The optimisation is performed through the gradient of the step-infidelity function $L(dw,\delta t)$, computed using a quantum computer.}
        \label{fig:sketch}
\end{centering}
\end{figure*}

\section{Method}
\label{sec:methods}

We address the simulation of a quantum system with Hamiltonian $\hat{H}$. 
For clarity of exposition we will assume that $\hat{H}$ is time independent, but it is not a requirement of the algorithm. 
The time evolution operator for a small time-step $\delta t \in \mathbf{R}$ is  $e^{-i\hat{H} \delta t}$.
Let $|\psi_{w(t)} \rangle$ be the parameterized ansatz state approximating the exact quantum state of the system $|\Psi (t)\rangle $ at time $t$, where $w(t) \in \mathbf{R}^{p}$ is the vector of its $p-$parameters. We assume $|\psi_{w(t)} \rangle $ is a unit vector and define the evolved state $|\phi(t+\delta t) \rangle = e^{-i\hat{H}\delta t}|\psi_{w(t)} \rangle$. 
From now on, we will indicate $w(t)$ as $w$ to simplify the notation, implying that the parameters we are referring to are the one assigned to the ansatz at time $t$, except when explicitly indicated. 
Also, we will shorten $|\phi(t+\delta t) \rangle $ to $|\phi(\delta t) \rangle$.

In order to variationally approximate the time evolution of the system, we aim to maximize the overlap between the evolved parameterized state $|\phi(\delta t) \rangle$ and the state $|\psi_{w+dw} \rangle$, where $dw \in \mathbf{R}^{p}$ is a vector of parameter variations. 
We want to find $dw$ such that 

\begin{equation}
\label{eq:argmax}
    \argmax_{dw \in \mathbb{R}^{p}} | \langle \phi(\delta t) | \psi_{w+dw} \rangle |^{2} \quad .
\end{equation}

This intermediate overlap optimization is routinely used in classical simulations of the dynamics of quantum systems, and its application to quantum circuits has been recently suggested in the context of quantum tensor networks \cite{lin2020real,Barratt_2021} and symmetry-preserving variational wavefunction \cite{otten2019noiseresilient}.
Similarly to \cite{Stokes_2020}, we derive the condition indicated in Eq. (\ref{eq:argmax}) defining the projected real-time evolution, as can be seen in Appendix \ref{appendix:proj_rte}. 
The optimal update $dw^{\star}$ therefore minimizes the step-infidelity

\begin{align}
\label{eq:cost_function}
    L(dw,\delta t) = \frac{1- | \langle \phi(\delta t) | \psi_{w+dw} \rangle |^{2}}{\delta t^{2}},
\end{align}

where the $\frac{1}{\delta t ^{2}}$ factor has been added to make $L$ independent from the time step size in the limit of $\delta t \rightarrow 0$. 
More details about the introduction of this factor can be found in  Appendix \ref{appendix:dt_exp}.
Given that the wave-function is encoded as the circuit $C(w)$, substituting  into the loss function in Eq. (\ref{eq:cost_function}) the definition $|\psi_w \rangle = C(w) |0 \rangle $ we obtain

\begin{equation}
   L(dw,\delta t) = \frac{1- |\langle 0 | C^{\dagger}(w) e^{i\hat{H} \delta t} C(w + dw) |0 \rangle |^{2}}{\delta t^{2}},
\end{equation}
where the second term on the right-hand side is an expectation value which can be sampled on a quantum computer. 
Another method that can be considered to measure the overlap between $\ket{\psi_{w+dw}}$ and $\ket{\phi(\delta t)}$ is the SWAP test \cite{Gottesman_2001,Buhrman_2001}, although it requires ancillary qubits and long range gates.
For these reasons, we will not consider this method in the following discussion.
The time evolution operator $e^{-i\hat{H} \delta t}$ is encoded in the form of a Trotter-Suzuki decomposition \cite{Trotter1959,Suzuki1991}.

The optimal $dw^\star$ are determined by iteratively descending along the steepest direction given by the gradient $\frac{\partial}{\partial dw_i}  L(dw, \delta t)$, starting from an initial guess $dw_{0}$.   
In general, the gradient of a quantum expectation value involving parameterized quantum circuits can be determined using finite differences \cite{schuld2018der} or simultaneous perturbation techniques \cite{Spall1998}.
However, those techniques rely on an approximation of the real gradient.
To improve the accuracy, we consider circuits of the general form
\begin{equation}
    C(w) = V_{p}U_{p}(w_{p})V_{p-1} \dots V_{1} U_{1}(w_{1})
\end{equation}
where the gates $V_k$ do not depend on any parameter and the parameterized gates  $U_{j}(w_{j})$ are of the form
\begin{equation}
    U_{j}(w_{j}) = e^{-\frac{i}{2}w_{j} G_j } = \cos(w_{j}/2)\mathbb{I} -i \sin(w_{j}/2) G_j,
\end{equation}
with $G_{j}^{2}= \mathbb{I}$. 
We remark that in general $G_{j}$ can be any tensor product of Pauli operators.
In this case, the  gradient can be computed exactly in a hardware-friendly way using the parameter shift rule  \cite{schuld2018der,Mitarai_2018,parrish2019hybrid,crooks2019gradients,mari2020estimating,banchi2020measuring}, obtaining

\begin{multline}
    \frac{\partial}{\partial dw_i}  L(dw,\delta t) = \\ = \frac{L(dw+s e_{i},\delta t) - L(dw-s e_{i},\delta t) }{2 \sin{(s)}},
\end{multline}
where $e_i$ is the versor in the $i$-th direction and $s \in \mathbb{R}$. 
We remark that our p-VQD algorithm does not require an ancilla qubit to perform measurements.

The optimisation of $dw$ is then performed according to a standard gradient descent scheme

\begin{equation}
    dw^{\text{new}} = dw^{\text{old}} - \eta \nabla_{dw} L(dw^{\text{old}}, \delta t),
\end{equation}
with learning rate $\eta \in \mathbf{R}^{+}$. 
A single optimisation step requires $O(p)$ measurements. 
The optimisation continues until the loss function goes below the desired threshold $\nu$. 
Finally, once $dw^{\star}$ is determined, the parameters at time $t+\delta t$ are obtained by

\begin{equation}
    w(t+\delta t) = w(t) + dw^{\star}(t)
\end{equation}

We highlight that the circuit width and depth are the same from the beginning to the end of the simulation. A sketch of the algorithm is shown in Fig \ref{fig:sketch}, while an accurate study of the total hardware cost can be found in Section \ref{sec:results}. 

\subsection{Relationship with other methods}
In order to understand the connection of our approach with existing methods proposed in the quantum computing setting, it is conceptually interesting to explicitly take the limit of a vanishing time step. 
In this limit, the parameter updates obtained with p-VQD is $dw = \dot{w} \delta t$ and $\dot{w}$ is the solution of the equation 
\begin{multline}
\label{eq:McLachlan}
    \sum_{j} \text{Re}\big[ G_{kj} \big] \dot{w}_{j}  = \text{Im}\bigg[\langle \partial_k \psi_w|H|\psi_w \rangle \bigg] + \\ + i\langle \psi_w |H| \psi_w \rangle \langle \partial_k \psi_w | \psi_w \rangle
\end{multline}
where $G_{kj}$ is the Quantum Geometric Tensor (QGT) 
\begin{equation}
\label{eq:QGT}
    G_{kj}(w) = \bigg\langle \frac{\partial \psi_w}{\partial w_k} ,  \frac{\partial \psi_w}{\partial w_j} \bigg \rangle - \bigg \langle \frac{\partial \psi_w}{\partial w_k} ,  \psi_w \bigg \rangle \bigg \langle \psi_w , \frac{\partial \psi_w}{\partial w_j}  \bigg \rangle \quad .
\end{equation} 
This expression for $dw$ coincides with that given by the McLachlan's variational principle \cite{McLachlan1964,Yuan2019tv} (considering also a time-dependent global phase on the ansatz state). 
Both McLachlan's and the time-dependent variational principle suffice to simulate real time dynamics of closed systems.
Moreover, it is possible to prove that the two approaches are equivalent under certain assumptions \cite{Yuan2019tv}.
However, on the practical point of view, there are some differences between them due to numerical instabilities that we will analyse further in Sec. \ref{sec:results}.
Considering the expression for the QGT given in Eq. (\ref{eq:QGT}), the TDVA relies on the evaluation of its imaginary part, which is skew-symmetric, making its inversion unstable when the off-diagonal elements are close to $0$.
The McLachlan's variational principle, on the contrary, requires the real part, as can be seen in Eq. (\ref{eq:McLachlan}).
For this reason, it is suggested that McLachlan’s is the most consistent variational principle for quantum simulation \cite{Yuan2019tv}. 
A detailed derivation of Eq. (\ref{eq:McLachlan}) can be found in Appendix \ref{appendix:var_principle}.

Another method conceptually similar to pVQD is the Restarted Quantum Dynamics (RQD), independently proposed in a recent preprint \cite{otten2019noiseresilient}.
This technique also solves an optimisation problem to obtain the time-evolved state.
However, the RQD is based on the minimisation of the infidelity squared by varying the parameters of the ansatz state, which can lead to different optimisation dynamics.

\subsection{Barren plateaus}
It has recently been shown that the optimisation of shallow circuits can also be affected by so-called barren plateaus if the cost function is global \cite{Cerezo_2021}.
This covers the case of operators in the form of
\begin{equation}
    O = a \mathbf{I} + c O_{1} \otimes O_{2} \otimes \dots \otimes O_{i},
\end{equation}
when $O_{i}$ is a non-trivial projector ($O^{2}_{i} = O_{i} \neq \mathbf{I}$) acting on subsystem $S_i$.
The  operator $O_{G} =  \mathbf{I} - |0\rangle \langle 0|$  that we use to estimate the cost function in Eq. \ref{eq:cost_function} belongs to this class, meaning that our cost function is global.

Barren plateaus are usually recognized by measuring the variance of the components of the gradient \cite{Cerezo_2021,McClean_2018,haug2021optimal,Grant_2019,bravoprieto2020variational}.
However, it is possible to show that both the fidelity between two states and the variance of its gradient have a non-vanishing lower bound (in the limit of a large number of qubits) if the two states differ by an infinitesimal transformation \cite{haug2021optimal}.
Therefore, p-VQD avoids such plateaus entirely because at every time step we optimise the infidelity between the current state $\ket{\psi_{w(t)}} $ and the infinitesimally evolved state $U(\delta t)\ket{\psi_{w(t)}}$.

A more general approach to avoid barren plateaus consists in replacing the global cost function by one that is local but has the same minimum \cite{Cerezo_2021}.
The global operator $O_G$ (in our case, encoding the infidelity) can be replaced by its local counterpart
\begin{equation}
\label{eq:local_loss}
    O_{L} = \mathbf{I} - \frac{1}{N}\sum_{j=1}^{N} |0_j\rangle \langle 0_j| \otimes \mathbf{I}_{\bar{j}}
\end{equation}

where $\mathbf{I}_{\bar{j}}$ is the identity operator on all qubits except for $j$ . 
Since $\expval{O_{L}}_w = 0 \Leftrightarrow \expval{O_{G}}_w = 0$, in the limit of small $\delta t$ the generated dynamics will also be equivalent to the McLachlan's variational principle.
In Sec. \ref{sec:results} we discuss some numerical results comparing the two approaches.


\section{Results}
\label{sec:results}

To demonstrate a practical application of the p-VQD algorithm, we consider the Transverse Field Ising Model on an open chain, 

\begin{equation}
H= J \sum_{i=0}^{N-1} \sigma^{z}_{i}\sigma^{z}_{i+1} + h\sum_{i=0}^{N} \sigma^{x}_{i} \quad .
\end{equation}

The first term accounts for interaction between spins while the latter represents a local and uniform magnetic field along the transverse direction $x$. 
For our simulations, we considered $J=\frac{1}{4}$, $h=1$ and, except when explicitly indicated, $N=3$ spins. 
We compare the time-evolution obtained through the p-VQD against the more-estabilished TDV-Algorithm.
We analyse the ideal case of a simulation in which we have access to the state vector produced by the quantum circuit (state-vector simulation) and the case in which to gain information about the quantum state we have to repeatedly measure the qubits (multi-shot simulation). 
The two simulations coincide in the limit of infinite samples.
However, when the number of samples is finite, statistical fluctuations produce a noise on the results which we will refer to as shot noise.
For both state-vector and multi-shot simulations we use IBM’s open-source library for quantum computing, Qiskit \cite{Qiskit}.

We consider a circuit ansatz of the form
\begin{align}
\label{eq:ansatz}
    C(w) &= \prod_{l = 1}^{d} c_{l}(w_{l}) \\
    &= \prod_{l = 1}^{d} \bigg[ \prod_{i= 1}^{N} R_{\alpha}^{(i)}(w_{i,l}) \bigg] \bigg[ \prod_{j = 1}^{ N-1} e^{-i w_{j,l}\sigma^{z}_{j}\sigma^{z}_{j+1}} \bigg],
\end{align}

where $R_{\alpha}^{(i)}(w_{i,l}) = e^{-i w_{i,l} \sigma^{\alpha}_{i}}$ is a single qubit rotation around the $\alpha = \{x,y\}$ axis.
Every block $c_{l}(w_l)$ has a layer of single qubits rotations followed by a layer of entangling two-qubits gates.
The total number of blocks, or depth, is $d$.
When $\alpha = x$ , a block $c_{l}$ is equivalent to the Trotter-Suzuki approximation of the unitary operator $e^{-i\hat{H}\delta t}$.
A more general parametrization is obtained by alternating rotations around  the $x$ and $y$ axis ($\alpha = x$ iff $l$ is odd, $\alpha = y$ otherwise).
The representation power and the number of variational parameters can be increased by making the ansatz deeper.  
For the chosen system, the two ans\"{a}tze perform similarly when shot noise is neglected.
However, alternating blocks of $x$ and $y$ rotations proved to be more stable in presence of shot noise. 
In the following we will be studying the latter unless otherwise noted.

As a first comparison, we consider the integrated infidelity $\Delta_F(T)$ achieved by both algorithms with respect to the exact simulation of the system over an entire time evolution from $t=0$ to $t=T$.
This quantity can be expressed as

\begin{equation}
    \Delta_F(T) = \int_{0}^{T} \left( 1 - |\langle \Psi(t) | \psi_{w(t)} \rangle|^{2} \right) dt \quad .
\end{equation}

\begin{figure}
        \includegraphics[width=1.0\columnwidth]{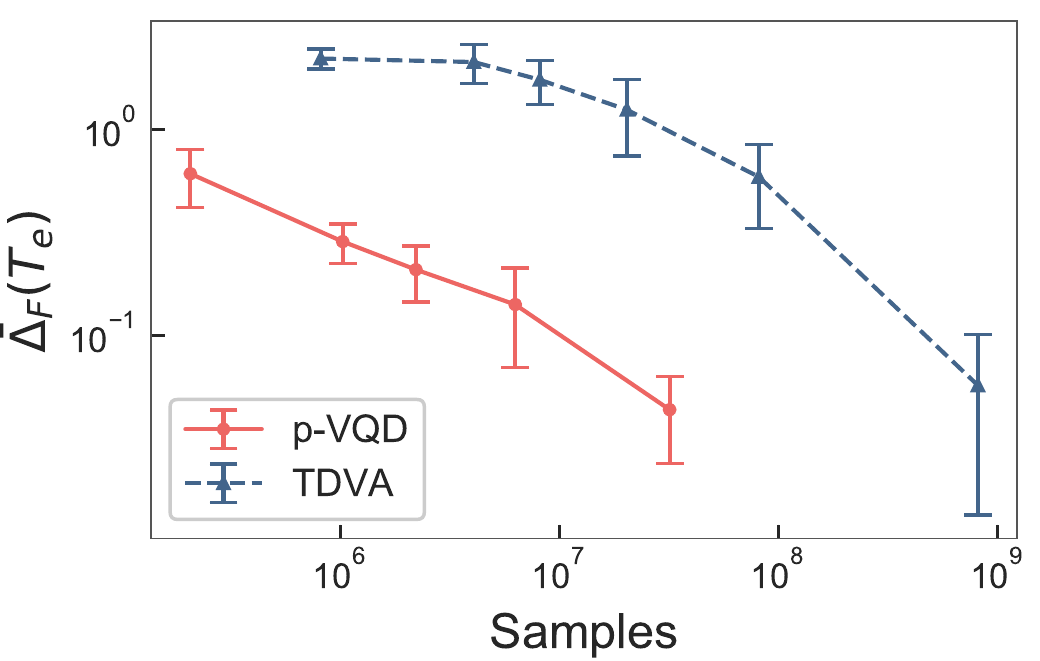} 
        \caption{Mean error on fidelity accumulated over an entire time evolution for the two different methods. The plot shows, as a function of the total samples required, the fidelity error accumulated by the algorithm over an entire time evolution. The total time of evolution is $T_{e}=3$, the number of time steps is $n_{\text{steps}} = 60$, the number of spins is $N=3$ and the ansatz chosen is the custom one with $d=3$ for both p-VQD and TDVA. For each number of samples the experiment has been iterated $n_{\text{iter}} = 10$ times and the data shown represent mean and standard deviation of those values.}
        \label{fig:fid_comp}
\end{figure}

We have performed several simulations both with p-VQD and the TDVA for different shots per circuit evaluation (total number of samples) and report the mean performance in Fig. \ref{fig:fid_comp}.

At a fixed  number of samples, the integrated infidelity $\Delta_{F}$ for the p-VQD is up to an order of magnitude below the TDVA one for the same number of time step. 
In Appendix \ref{appendix:var_principle} it is shown that with small time steps  and in the limit of infinite samples (ideal measures) the parameter variation found by p-VQD coincide with the solution of the Euler-Lagrange equation generated by the McLachlan's variational principle.
The matrices of the Euler-Lagrange equations often show high condition number \cite{Demmel1987}, in particular those generated by the time-dependent variational principle \cite{Yuan2019tv}.
These ill-conditioned matrices are likely to produce large variations on the results when subject to small variations of the coefficients, as those produced by shot noise. 
We remark that shot noise is unavoidable for quantum computers, even for fault-tolerant devices.
Improving iteratively the solution of the Euler-Lagrange equation leads to the estimate of the parameters' variation, but does not require matrix inversion.
As a result, we obtain a more stable algorithm against shot noise. 

\begin{figure}[!h]
        \includegraphics[width=1.0\columnwidth]{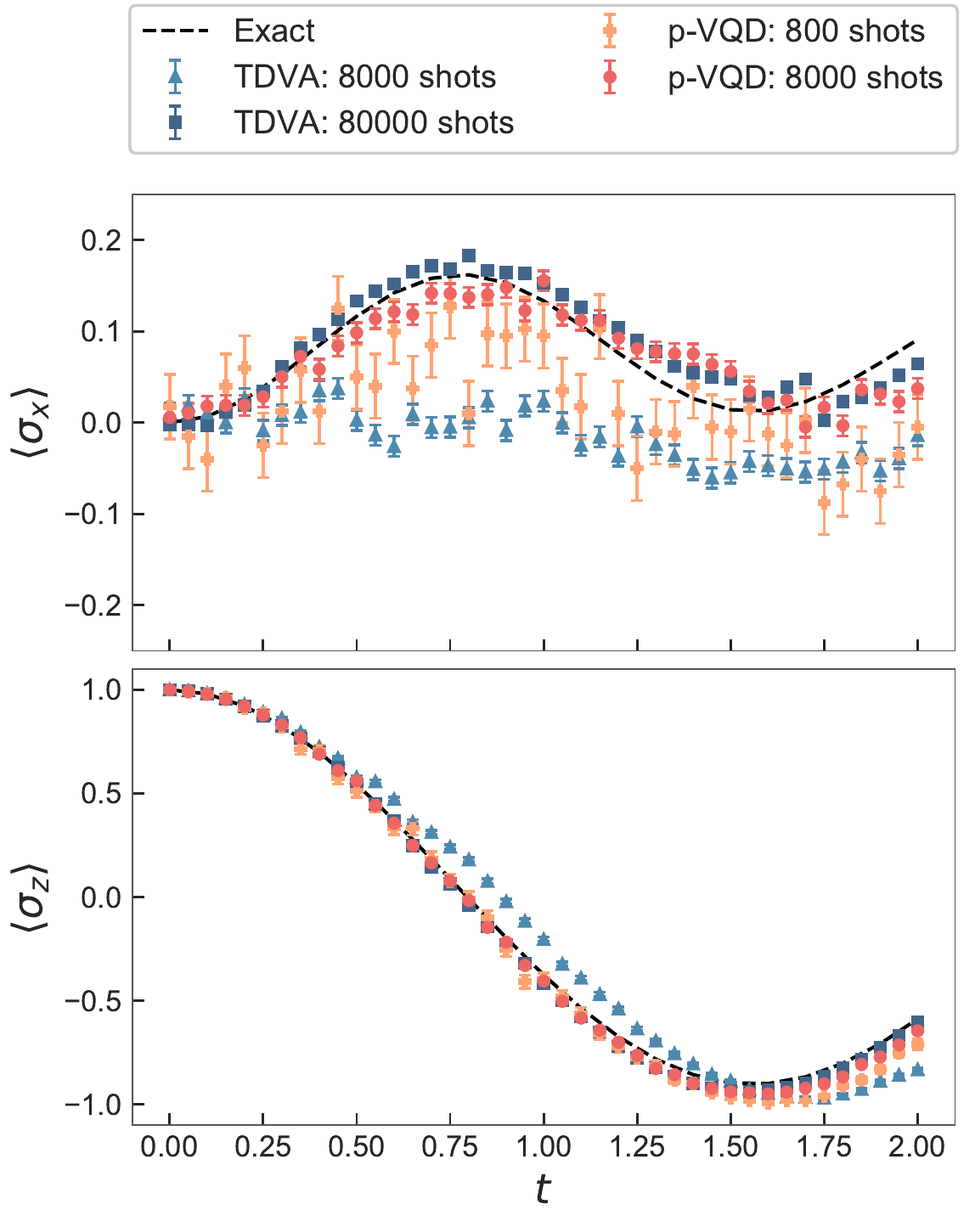} 
        \caption{Total magnetization measured on simulated time evolution. The plot shows the expectation values of Pauli operators for systems simulated with p-VQD and TDVA at different shot number. The magnetization is evaluated on a chain of $N=3$ spins and we used the custom ansatz with $d=3$ for both the algorithms. We indicated the number of shots for circuit evaluation instead of total samples: $800,8000$ and $80000$ shots are equivalent to $\sim 10^{6},10^{7},10^{8}$ total samples for the p-VQD and $\sim 10^{7},10^{8},10^{9}$ for the TDVA, respectively.}
        \label{fig:shot_comparison}
\end{figure}

In Fig. \ref{fig:shot_comparison} we compare the expectation values of the total magnetization along the $\hat{x}$ and $\hat{z}$ axis for the states evolved according to the two algorithms.
The results are reported for a different number of measurements per circuit. 

The magnetization along the $\hat{x}$ axis proved to be the most difficult to capture for both the algorithms, suggesting that the problem could be the ansatz choice.
We note that both the algorithms converge to the exact simulation values as the number of shots increases.
In general, p-VQD shows comparable results with TDVA using one order of magnitude fewer shots.

\begin{figure}
        \includegraphics[width=1.0\columnwidth]{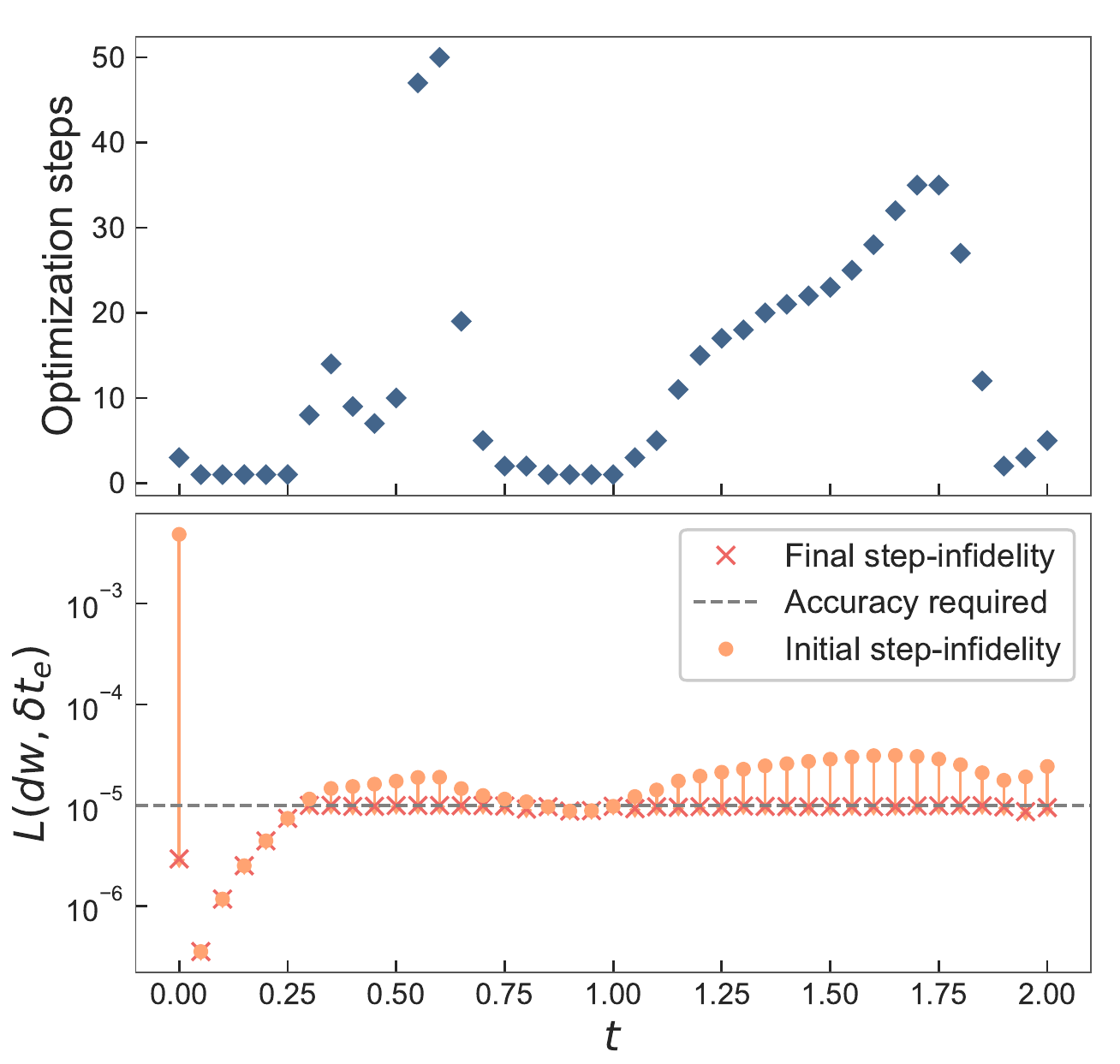} 
        \caption{Number of optimisation steps required (top) to decrease accuracy error per step (bottom). The algorithm optimizes the parameters shift $dw$ until the step-infidelity $L(dw,\delta t_{e})$ is below the threshold at $10^{-5}$. The time step chosen is $\delta t_{e}= 0.05$. The analysis is performed on a state-vector simulation of a system of $N= 3$ spins, using the custom ansatz with depth $d=3$.}
        \label{fig:fid}
\end{figure}

To further analyse the performance of our algorithm, we studied the behaviour of the step-infidelity $L(dw,\delta t)$. 
In Fig. \ref{fig:fid} we show the number of optimisation steps required as a function of time and the consequent decrease of the cost function.

We remark that choosing the previous $dw^{\star}$ as an initial guess often leads to convergence in only one iteration. 
More details about the convergence  of the step-infidelity in a single time step can be found in Appendix \ref{appendix:single_opt}.
    
We characterise the hardware cost of p-VQD when the number of variational parameters increases. 
Since the number of measures required by p-VQD scales linearly ($O(p)$) with the number of variational parameters, it has a lower asymptotic cost with respect to TDVA, which is quadratic.
We can provide an accurate estimation of the hardware cost of p-VQD as an upper-bound to the total number of samples needed per simulation.
We indicate with $n_s$ and $n_t$ the number of shots and time steps, respectively.
In the spirit of the iterative methods, we suppose that the procedure will find a solution $dw^{*}$ in at most $M$ steps. In this case, we have that the total number of hardware measurements is
    
\begin{equation}
    \label{eq:upper_bound_iter}
    N_{\text{samples}} \leq 2M n_{t} n_{s} p
\end{equation}

\begin{figure}
        \includegraphics[width=1.0\columnwidth]{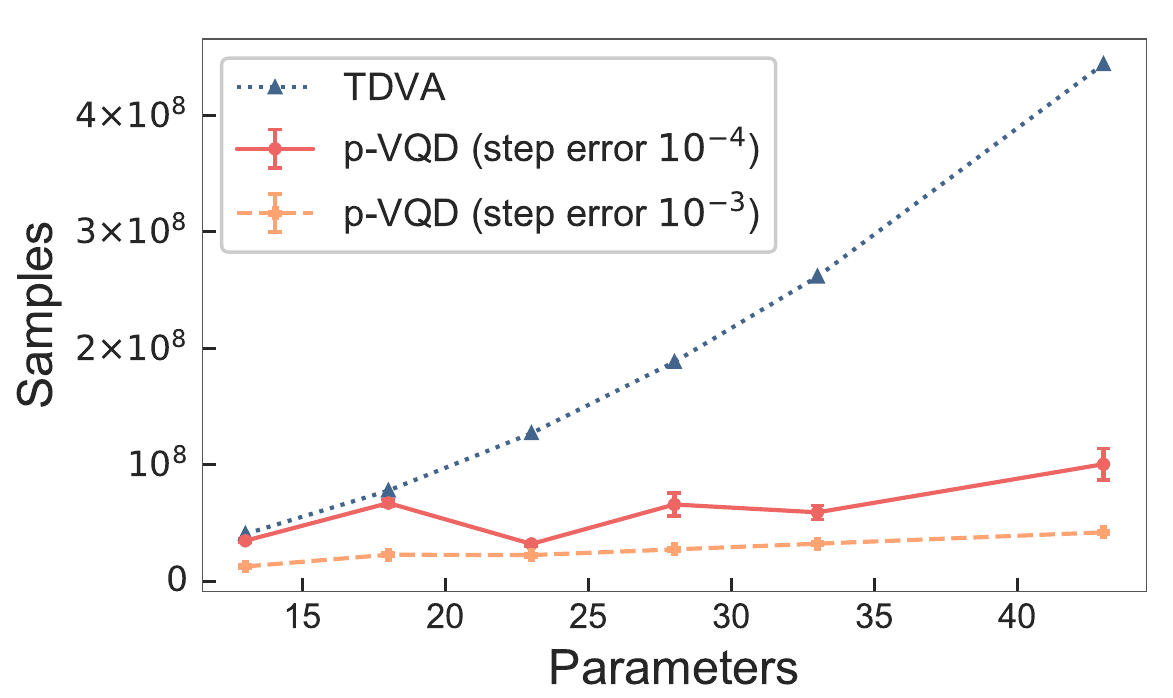}
        \caption{Measurements required as a function of circuit parameters. The plot shows the total number of circuits created and measured using p-VQD and TDVA. The number of required measurements by the TDVA is fixed, while in the p-VQD it depends on the optimisation of the parameter variation. We fixed the number of spins at $N=3$ and increased the depth of the custom ansatz from $d=2$ to $d=8$; the parameters varied accordingly. We considered $n_{\text{shots}}=8000$ per circuit evaluation for both methods and performed multiple simulations for p-VQD, showing mean values and standard deviations of the results. Different error on the accuracy per step are showed for p-VQD.}
        \label{fig:depth}
\end{figure}

Some terms in Eq \ref{eq:upper_bound_iter} are not free parameters: the number of shots $n_s$ depends on the accuracy required, while the number $M$ of optimisation steps depends also on the ansatz and on the Hamiltonian of the system considered. 
More details about the role of $n_s$ can be found in the Appendix \ref{appendix:shot_accuracy}.
To characterise the dependence of $M$ upon the number of parameters $p$  we performed different simulations increasing the depth of our ansatz and reported the results in Fig \ref{fig:depth}.
From the figure we deduce that  $M \sim  O(1)$, therefore it does not scale up by increasing the number of parameters in the ansatz circuit.
Under these circumstances, the total dependence of the computational cost of the algorithm upon the number of parameters is indeed $O(p)$.

All the simulations have been done fixing a number of shots $n_{\text{shots}}$ per circuit evaluation. For p-VQD, we made multiple simulations and then reported the mean values of our results. For TDVA the number of circuit evaluation is fixed and can be estimated a priori. This comparison does not consider the fidelities of the results obtained with the exact simulation, we know from what we showed in Fig. \ref{fig:fid_comp} that p-VQD has a lower error when the total number of samples is comparable with TDVA. We note that the number of samples required scales approximately linearly with the number of parameters, with fluctuations due to different optimisation steps required. The lower the step-infidelity required, the higher will be the number of shots and possibly of the optimisation steps, resulting in greater fluctuations on the total number of samples.
In this case, we remark that more advanced methods like the use of an adaptive learning rate $\eta$ can be used to improve convergence performance of p-VQD.

\begin{figure}[!h]
        \includegraphics[width=1.0\columnwidth]{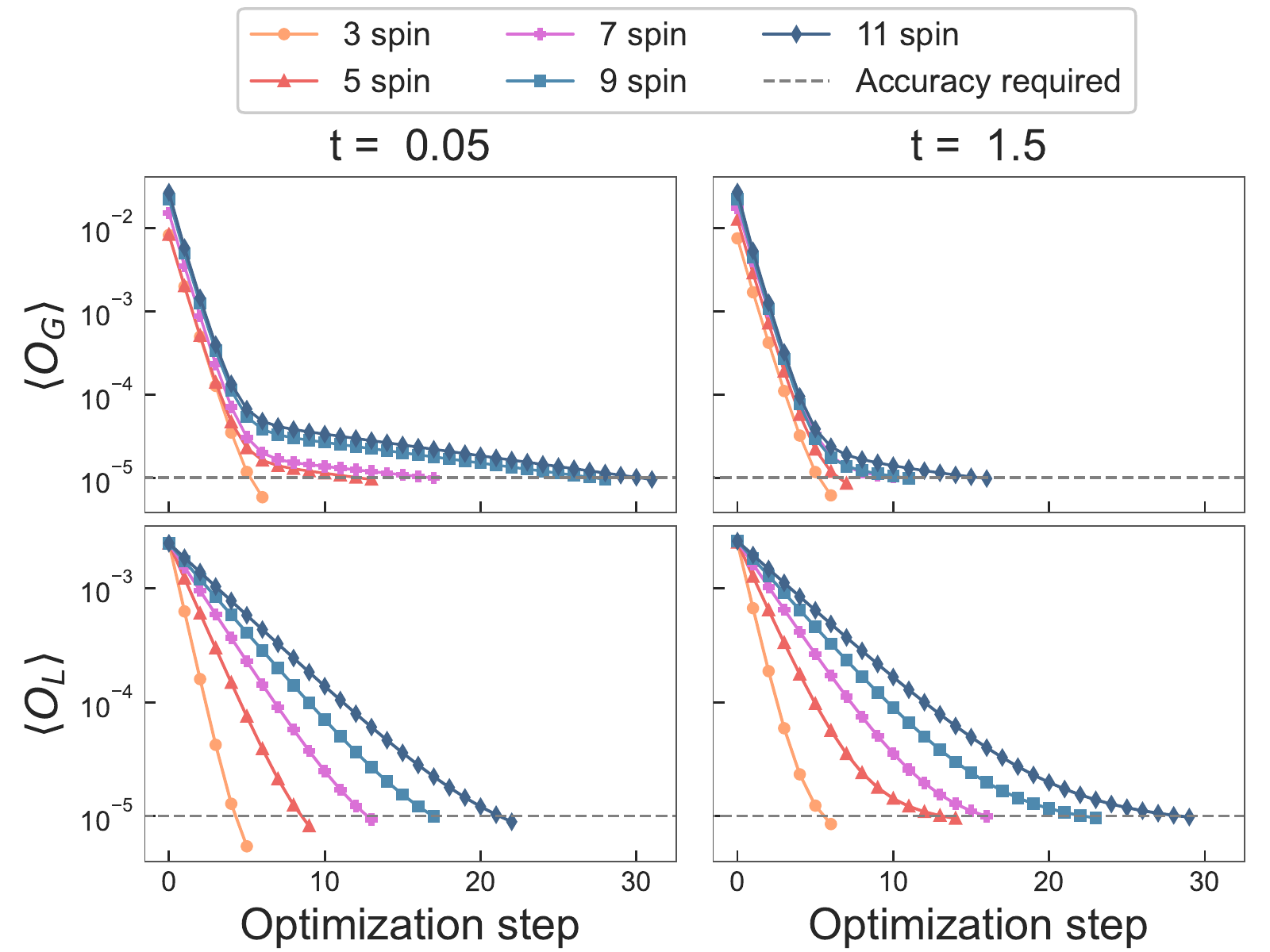}
        \caption{Optimisation steps required per time step when increasing the qubits number. The plot compares the use of the global and the local cost function in the optimisation process of two different time steps. The analysis is performed using the state-vector simulator and an accuracy per step of $10^{-5}$. The ansatz has $\alpha=x$ at every layer and $d=3$. The number of parameters varies accordingly. The two cost functions show similar performances in the number of iterations required.}
        \label{fig:cost_fun_comparison}
\end{figure}

Finally, we compare numerically the performance of a global and a local cost function. 
As discussed in Sec. \ref{sec:methods}, both $O_{G}$ (Eq. \ref{eq:cost_function}) and $O_{L}$ ( Eq. \ref{eq:local_loss}) can be used.
In Fig. \ref{fig:cost_fun_comparison} we show the minimisation of both cost functions for different time steps and increasing number of spins $N \in [3,11]$, obtained using the exact state-vector simulator.

All the curves are obtained by setting the initial condition $dw_{0}= 0$.
We highlight that our numerical findings point to an efficient scaling with the system size.
We remark that the first optimisation steps of $O_G$ depend very weakly on the number of qubits.
However, we also note that in general barren plateaus during optimisation may not show up until even larger systems are considered \cite{Cerezo_2021,bravoprieto2020variational}.
For this reason, one should benchmark which cost function behaves best for their specific system and size.

\section{Discussion}
\label{sec:discussion}

In this work, we presented a new algorithm for the efficient variational simulation of the real-time evolution of quantum systems. We have shown that it is asymptotically more hardware efficient than the time-dependent variational algorithm (TDVA), while retaining an higher accuracy. 
Considering the projected time evolution, we avoid numerical instabilities due to the matrix inversion combined with statistical fluctuations due to finite shot measurements.
We have numerically investigated the absence of barren plateaus, paving the way towards the simulation of larger quantum systems.
One possible application of our approach is to study the dynamical properties of two-dimensional interacting systems, a notoriously difficult problem in classical computational physics.
Similarly to all other variational algorithms, the choice of the right parametrization is fundamental for the algorithm to succeed. In this sense, having an efficient quantum algorithm to perform variational time evolution is essential 
to compare to classical results obtained with variational states either based on tensor networks \cite{vidal_efficient_2004,daley_time-dependent_2004,white_real-time_2004}, or neural networks \cite{carleo_solving_2017,schmitt_quantum_2020}. 
A possible improvement to further enhance the efficiency of our approach concerns the estimation of the gradient. 
At present, a drawback of our method is that the circuit constructed on the quantum device is approximatively twice as deep as the ansatz used to represent the system. 
However, by suitably controlling the number of two-qubits gates in the chosen ansatz, representing the major source of circuit error, we believe that p-VQD can already be used to simulate small quantum systems on available devices.

\section*{Data availability}
The code used to run the simulations in this article has been written in Python using Qiskit \cite{Qiskit}. 
It is open source and can be found on GitHub \cite{barison2021github}.

\section*{Acknowledgments}
We thank T. Haug and A. Green for fruitful discussions and  J. Gacon for constructive remarks on our code.
We also thank one anonymous referee for suggesting the possibility of using the local cost function, Eq. \ref{eq:local_loss}.

\printbibliography



\appendix
\onecolumn

\section{The projected real time evolution}
\label{appendix:proj_rte}

We aim to simulate the evolution of a quantum system by acting on the parameters of the variational ansatz that approximates the real state of the system. The parameter variation has to satisfy Eq. (\ref{eq:argmax}). We are going to see how this condition can be derived defining the projected real-time evolution. Consider the projector $P_{w} = |\psi_{w} \rangle \langle \psi_{w} |$ and define the squared distance between the evolved parameterized state and its projection on the subspace spanned by $|\psi_{w+dw} \rangle$

\begin{equation}
    || |\phi(\delta t) \rangle - P_{w+dw} |\phi(\delta t)\rangle  ||^{2} \quad .
\end{equation}

To find the best approximation for the evolved state, we want to minimize this distance finding the optimal $dw$. Imposing this condition we obtain

\begin{align}
\begin{split}
    \argmin_{dw \in \mathbb{R}^{p}}|| |\phi(\delta t) \rangle - P_{w+dw} |\phi(\delta t)\rangle  ||^{2}  & =  \argmin_{dw \in \mathbb{R}^{p}} \bigg( \langle \phi(\delta t)| - \langle \phi(\delta t)|P_{w+dw} \bigg) \bigg( |\phi(\delta t) \rangle - P_{w+dw} | \phi(\delta t) \rangle \bigg) \\
    & = \argmin_{dw \in \mathbb{R}^{p}} \bigg[ 1 - 2\langle \phi(\delta t)| P_{w+dw} | \phi(\delta t) \rangle + \langle \phi(\delta t)| P^{2}_{w+dw} | \phi(\delta t) \rangle \bigg] \\ 
    &= \argmin_{dw \in \mathbb{R}^{p}} \bigg[ 1 - \langle \phi(\delta t)| P_{w+dw} | \phi(\delta t) \rangle  \bigg] \\ 
    & = \argmax_{dw \in \mathbb{R}^{p}} | \langle \phi(\delta t) | \psi_{w+dw} \rangle |^{2}
\end{split}
\end{align}

where we used the idempotence property of the projector $P^{2}_{w+dw}= P_{w+dw}$. With the final equivalence we see that this condition is equivalent to Eq. (\ref{eq:argmax}).


\section{Relationship of step-infidelity with time step and parameter variation}
\label{appendix:dt_exp}

In this Appendix we explain the introduction of the factor $\frac{1}{\delta t ^{2}}$ in Eq. (\ref{eq:cost_function}).
Consider the overlap contained in the step-infidelity definition in Eq. (\ref{eq:cost_function}) and Taylor expand it in $\delta t $ and $dw$. First, expand to the second order the time evolution operator $e^{-iH\delta t} \sim \mathbf{I} - i H \delta t - \frac{H^2}{2}(\delta t)^2 $ to get

\begin{multline}
    \langle \psi_{w+dw} |(\mathbf{I} -i H \delta t - \frac{H^2}{2}(\delta t)^2) |\psi_w \rangle \langle \psi_w | (\mathbf{I} +i H \delta t - \frac{H^2}{2}\delta t^2) |\psi_{w+dw} \rangle =  \\  = |\langle \psi_{w+dw}| \psi_w \rangle |^{2}  -i \delta t \langle \psi_{w+dw} | H |\psi_w \rangle \langle \psi_w  |\psi_{w+dw} \rangle  + i\delta t \langle \psi_{w+dw} |\psi_{w} \rangle \langle \psi_w | H  |\psi_{w+dw} \rangle + (\delta t)^{2} |\langle \psi_{w+dw} |H| \psi_w \rangle |^{2} - \\  - \frac{\delta t^2}{2} \langle \psi_{w +dw} |H^2 | \psi_w \rangle \langle \psi_w | \psi_{w+dw} \rangle - \frac{\delta t^2}{2} \langle \psi_{w +dw} | \psi_w \rangle \langle \psi_w |H^2 | \psi_{w+dw} \rangle 
\end{multline}

Now we expand to first order in $dw$, obtaining $|\psi_{w+dw} \rangle = |\psi_w\rangle + \sum_{j} dw_{j} |\partial_j \psi_w \rangle$. The first order contribution in $\delta t$ vanishes, but also the first order contribution in $dw$, then we have

\begin{equation}
    \sum_{j} dw_{j} \left[ \langle \partial_j \psi_w | \psi_w \rangle - \langle \partial_j \psi_w | \psi_w \rangle  \right] = 0  \quad \quad \forall dw_j
\end{equation}

The final result reads

\begin{multline}
    |\langle \psi_{w+dw} | e^{-i H \delta t}| \psi_w \rangle |^{2}   \sim 1 + \delta t^{2} \bigg[ \sum_{j} \frac{dw_{j}}{\delta t} [2i \langle H \rangle_{w} \langle \partial_{j} \psi_w |\psi_w \rangle   - 2 \text{Im}[\langle \psi_w|H| \partial_{j}\psi_w \rangle ]]  - \\ - \text{Var}_{w}(H)   + \sum_{j,k} \frac{dw_{j}}{\delta t} \frac{dw_{k}}{\delta t} \langle \partial_{j} \psi_w | \psi_w \rangle \langle \psi_w | \partial_k \psi_w \rangle \bigg] \quad .
\end{multline}

As $\delta t \rightarrow 0$, also $dw_j \rightarrow 0 \quad \forall j$ and their ratio remains constant. 
Therefore, the addition of $\frac{1}{\delta t ^{2}}$ factor in Eq. (\ref{eq:cost_function}) makes it independent of the time step size in the limit of $\delta t \rightarrow 0$.


\section{Equivalence between the p-VQD and the McLachlan's variational principle}
\label{appendix:var_principle}

In this Appendix, we show that in the limit of small time step the parameter variation that fulfils the request in Eq. (\ref{eq:argmax}) is the same obtained by applying the McLachlan's variational principle \cite{McLachlan1964,Yuan2019tv}.

We start Taylor expanding the overlap to the second order: using the substitution

\begin{equation}
    |\psi_{w+dw} \rangle = |\psi_w\rangle + \sum_{k} dw_{k} |\partial_k \psi_w \rangle + \frac{1}{2}\sum_{k,j} dw_{k}dw_{j} |\partial_k \partial_j \psi_w \rangle  + o(dw^{3})
\end{equation}

we obtain

\begin{multline}
\label{eq:first_expansion}
     |\langle \phi(\delta t) | \psi_{w+dw} \rangle |^{2} =  |\langle \phi(\delta t) | \psi_{w} \rangle |^{2} + \sum_{k} \bigg[ \langle \phi(\delta t) | \psi_{w} \rangle \langle \partial_k \psi_{w} | \phi(\delta t) \rangle  + \langle \phi(\delta t) | \partial_k \psi_{w} \rangle \langle \psi_{w} | \phi(\delta t) \rangle \bigg] dw_k + \\
     + \sum_{k,j} \bigg[ \langle \phi(\delta t) | \partial_k \psi_{w} \rangle \langle \partial_j \psi_{w} | \phi (\delta t) \rangle + \frac{1}{2}\langle \phi(\delta t)| \partial_k \partial_j \psi_{w} \rangle \langle \psi_{w} | \phi(\delta t)\rangle + \langle \phi(\delta t) |  \psi_{w} \rangle \langle \partial_k \partial_j \psi_{w} | \phi(\delta t) \rangle \bigg] dw_k dw_j +o(dw^3)
\end{multline}

where we used $ |\phi(\delta t) \rangle = e^{-i H \delta t}|\psi_w \rangle $ as in the main text.

Then, we expand  the time evolution operator to the first order $ e^{-i H \delta t} = \mathbf{I} -iH\delta t +o(\delta t)$ and partially differentiate both sides of the normalization condition $||\psi_{w}||^{2} =1$ with respect to parameters $w_i$ and $w_j$ to obtain the two important properties

\begin{align}
    \langle \psi_w | \partial_k \psi_w \rangle & = - \langle \partial_k \psi_w |  \psi_w \rangle \\
    \langle \psi_w | \partial_k \partial_j  \psi_w \rangle +  \langle \partial_k \partial_j \psi_w |  \psi_w \rangle & = -  \langle \partial_k \psi_w | \partial_j \psi_w \rangle - \langle \partial_j \psi_w | \partial_k \psi_w \rangle \quad .
\end{align}

Substituting  in Eq. (\ref{eq:first_expansion}) we have

\begin{multline}
    |\langle \phi(\delta t) | \psi_{w+dw} \rangle |^{2} = |\langle \phi(\delta t) | \psi_{w} \rangle |^{2} + \\ 
    + i \sum_{k} \bigg[ \langle \psi_w |H| \psi_w \rangle \langle \partial_k \psi_w | \psi_w \rangle - \langle \psi_w |H| \psi_w \rangle \langle  \psi_w | \partial_k \psi_w \rangle + \langle \partial_k \psi_w | H | \psi_w \rangle - \langle  \psi_w | H | \partial_k \psi_w \rangle \bigg] dw_k \delta t + \\ 
    + \frac{1}{2}\sum_{k,j} \bigg[  2\langle \psi_w | \partial_k \psi_{w} \rangle \langle \partial_j \psi_{w} | \psi_w \rangle - \langle \partial_k \psi_w | \partial_j \psi_w \rangle - \langle \partial_j \psi_w | \partial_k \psi_w \rangle \bigg] dw_k dw_j
\end{multline}

where we neglected the third order contribution in $dw$ and $\delta t$. We notice that the second order term in $dw_k dw_j$ is the real part of the Quantum Geometric Tensor (QGT) , as expressed in Eq. (\ref{eq:QGT}). Finally, in the p-VQD algorithm we aim to find the $dw$ that maximizes the overlap in Eq.(\ref{eq:argmax}) , thus we impose the first order optimality condition \begin{equation}
\frac{\partial}{\partial dw_k} |\langle \phi(\delta t) | \psi_{w+dw} \rangle |^{2} = 0 \quad \forall k
\end{equation}

that in the limit $\delta t \rightarrow 0 $ gives the equation

\begin{equation}
    \sum_{j} \text{Re}\big[ G_{kj} \big] \dot{w}_{j}  = \text{Im}\bigg[\langle \partial_k \psi_w|H|\psi_w \rangle \bigg] +i\langle \psi_w |H| \psi_w \rangle \langle \partial_k \psi_w | \psi_w \rangle
\end{equation}

where we indicated with $G$ the QGT, as in the main text. This is the same evolution equation for parameters $w$ that is obtained when the McLachlan's variational principle is used to simulate real time dynamics of  closed systems with pure quantum states. For an extensive review of the variational principles for time evolution of pure and mixed states, see \cite{Yuan2019tv}.

\section{Optimisation routine for a single time-step}
\label{appendix:single_opt}

In this Appendix, we further analyse the optimisation routine of our algorithm. This optimisation is performed on the parameter variation $dw$ for every time step. In this case we focus on a state-vector simulation of the Transverse Field Ising Model on an open chain (see main text), considering only a single time step (without loss of generality, the third time step of the simulation).

\begin{figure}[!ht]
\centering
\includegraphics[width=0.5\columnwidth]{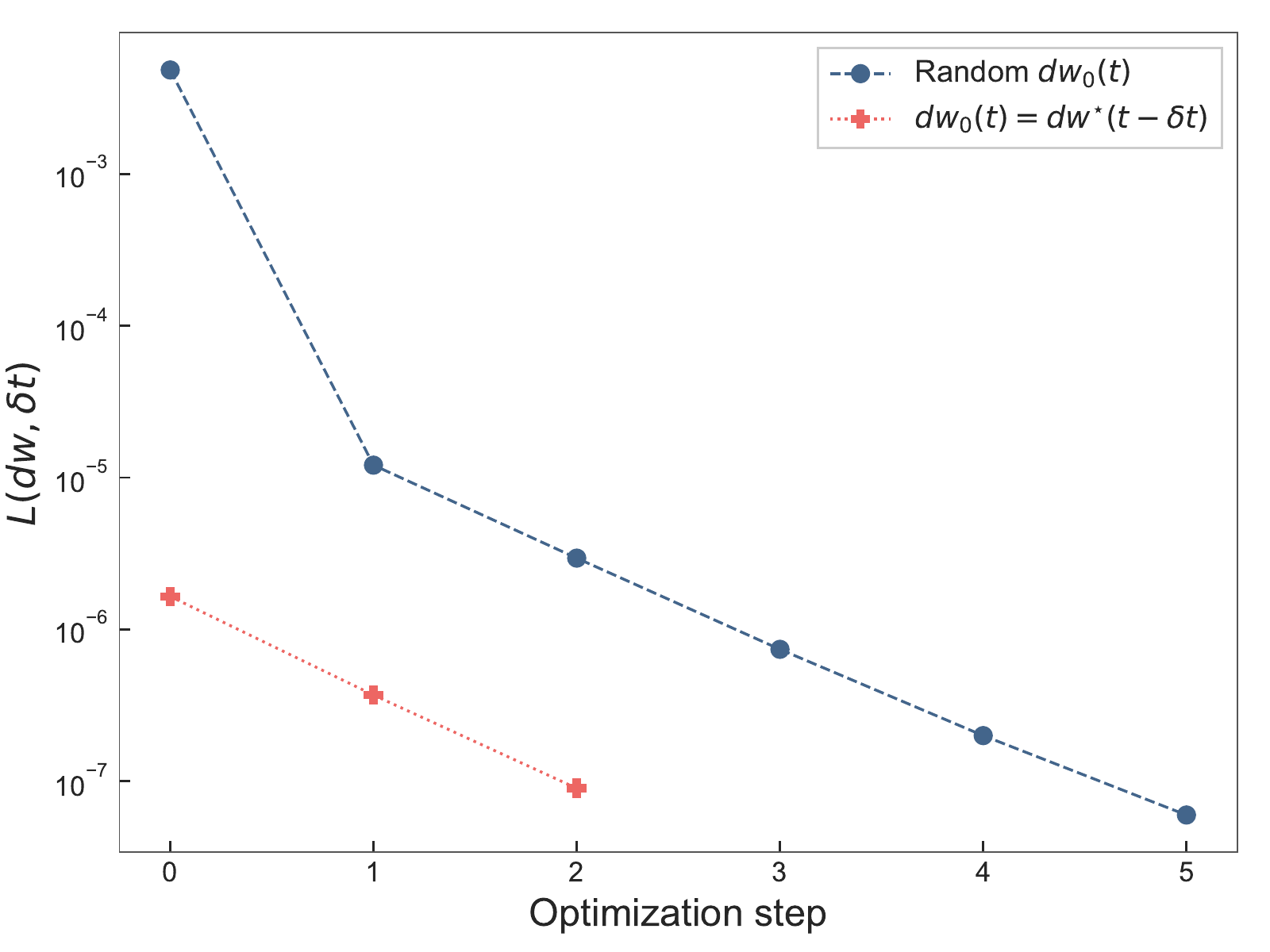} 
\caption{Infidelity as a function of the optimisation step. The optimisation procedure requires only a few optimisation steps to greatly decrease the step-infidelity. Moreover, we compare the random choice of an initial $dw_{0}$ of order $O(\delta t)$ to the educated guess $dw_{0} = dw^{\star}(t-\delta t)$. }
\label{fig:opt_step}
\end{figure}

As Fig. \ref{fig:opt_step} shows, at every optimisation steps the step-infidelity is reduced of nearly a order of magnitude. The initial choice $dw_{0} = dw^{\star}(t-\delta t)$ is shown to  reduce the number of steps required.

\section{Relationship between step infidelity and number of shots}
\label{appendix:shot_accuracy}

In this Appendix, we investigate the relationship between the step infidelity to reach in a p-VQD step and the number of shots $n_s$ required.
Given that the fidelity is equivalent to the probability of measuring a string of $0$s, with $n_s$ shots the minimum non-zero infidelity we can resolve is $\frac{1}{n_s}$. 
However, we stress that since we are optimising the infidelity by measuring the gradient, it is possible to converge to a state with a smaller step infidelity than this bound, depending on the variance of the gradient and on the optimisation algorithm employed.

\begin{figure}[!ht]
\centering
\includegraphics[width=0.50\columnwidth]{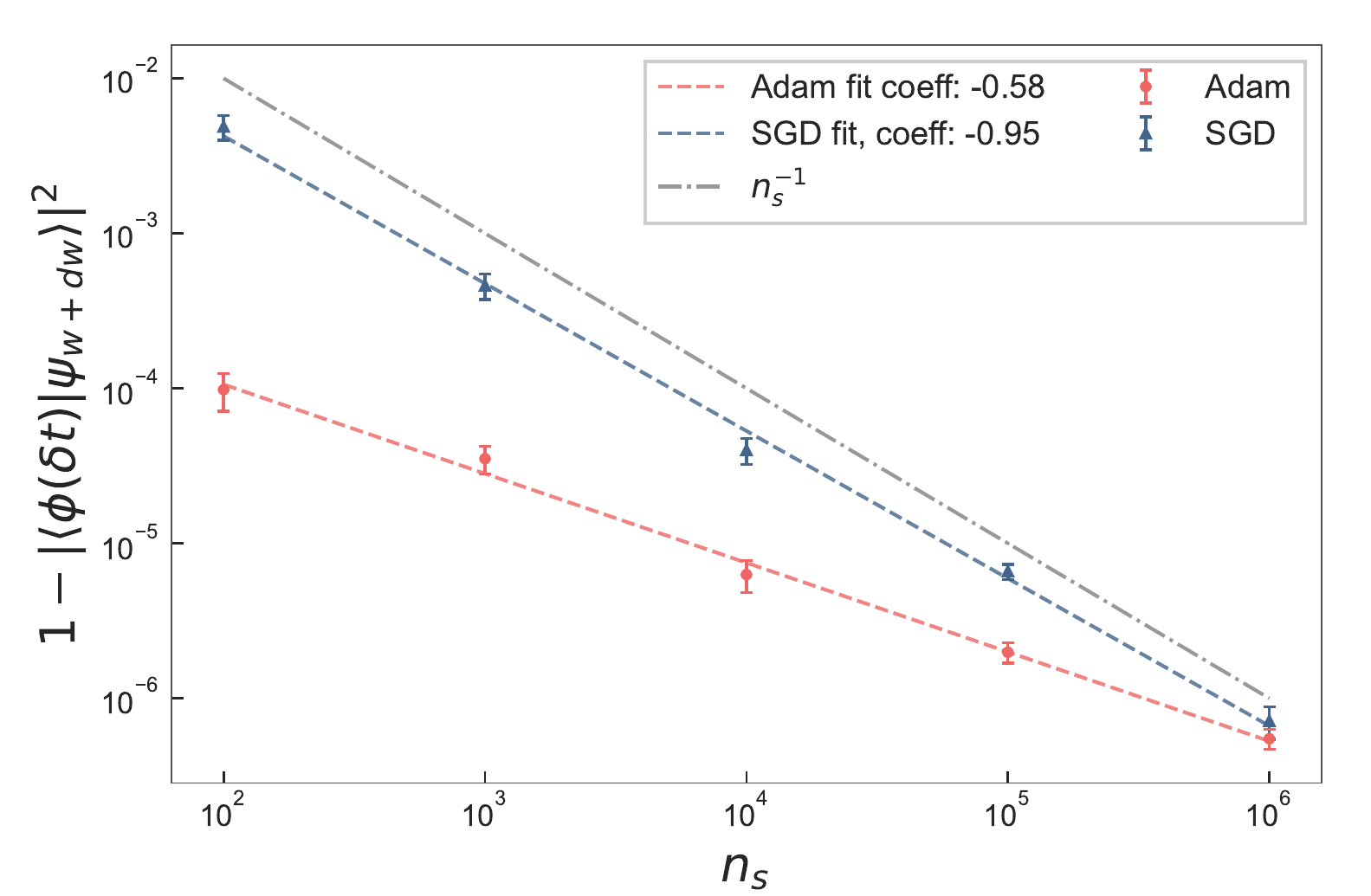} 
\caption{Asymptotic infidelity reachable using shots to sample the gradient. This plots illustrates the infidelity between the target and the ansatz state as the number of shots to sample the gradient increases.  The values reported are the mean and the standard deviations of multiple runs made both with the Stochastic Gradient Descent and Adam as optimizers. The dashed lines are obtained fitting the function  $f = kn_{s}^{\gamma} $ . The grey dashdotted line indicates the expected behaviour $n_{s}^{-1}$. In this simulation we considered $N=3$, $d=3$ and $M=150$.} 
\label{fig:shot_accuracy}
\end{figure}

To analyse the effects of the shots on the infidelity optimisation we considered a single time step of the algorithm and optimised the ansatz parameters using a gradient sampled with $n_s$ shots, while measuring the infidelity exactly using the state-vector.
For each number of shots $n_s$, we optimised the step infidelity and measured its convergence value. 
Due to the presence of shot noise, we repeated the optimisation multiple times and plotted the mean and the standard deviation of those calculations in Fig. \ref{fig:shot_accuracy}.
We performed the experiment using two different optimizers, Stochastic Gradient Descent and Adam, to study how this choice affects the infidelity minimisation.
We also fit the points to a function $f = kn_{s}^{\gamma} $ with $k$ and $\gamma$ free parameters,to estimate the exponent  $\gamma$. 
The expected $n_{s}^{-1}$ dependence is shown as a visual guide for clarity.

We can see that, when a small number of shots  is considered, the Adam optimizer is able to minimise the infidelity up to two order of magnitude below the minimum resolvable with shots.
When the number of shots increases, the performances of the two optimisers become comparable.

\end{document}